\documentclass[]{spie}  
\usepackage[]{graphicx}

\title{GPI PSF subtraction with TLOCI: the next evolution in exoplanet/disk high-contrast imaging} 

\author{Christian Marois\supit{a}, Carlos Correia\supit{a,b}, Rapha\"{e}l Galicher\supit{a,c}, Patrick Ingraham\supit{d}, Bruce Macintosh\supit{d}, Thayne Currie\supit{e}, Rob De Rosa\supit{f}
\skiplinehalf
\supit{a} National Research Council of Canada, 5071 West Saanich Rd, Victoria, V9E 2E7, Canada;\\
\supit{b} Centre for Astrophysics of the University of Porto, Rua das Estrelas 4150-762 Porto, Portugal;\\
\supit{c} LESIA, Observatoire de Paris, CNRS, UPMC Paris 6 and Denis Diderot Paris 7, 92195 Meudon, France;\\
\supit{d} Stanford University, 450 Serra Mall, Stanford, CA, United States;\\
\supit{e} University of Toronto, 50 St. George Street Room 101, Toronto, Ontario, Canada M5S 3H4;\\
\supit{f} School of Earth and Space Exploration, Arizona State University, PO Box 871404, Tempe, AZ 85287, USA
}

\authorinfo{Further author information: (Send correspondence to C. Marois)\\C. Marois: E-mail: christian.marois@nrc-cnrc.gc.ca, Telephone: 1 250 363 0023\\}

\begin{document} 
  \maketitle 
\begin{abstract}
To directly image exoplanets and faint circumstellar disks, the noisy stellar halo must be suppressed to a high level. To achieve this feat, the angular differential imaging observing technique and the least-squares Locally Optimized Combination of Images (LOCI) algorithm have now become the standard in single band direct imaging observations and data reduction. With the development and commissioning of new high-order high-contrast adaptive optics equipped with integral field units, the image subtraction algorithm needs to be modified to allow the optimal use of polychromatic images, field-rotated images and archival data. A new algorithm, TLOCI (for Template LOCI), is designed to achieve this task by maximizing a companion signal-to-noise ratio instead of simply minimizing the noise as in the original LOCI algorithm. The TLOCI technique uses an input spectrum and template Point Spread Functions (PSFs, generated from unocculted and unsaturated stellar images) to optimize the reference image least-squares coefficients to minimize the planet self-subtraction, thus maximizing its throughput per wavelength, while simultaneously providing a maximum suppression of the speckle noise. The new algorithm has been developed using on-sky GPI data and has achieved impressive contrast. This paper presents the TLOCI algorithm, on-sky performance, and will discuss the challenges in recovering the planet spectrum with high fidelity.
\end{abstract}

\section{Introduction}
\label{sec:intro}  

While the direct imaging field has progressed more slowly compared to radial velocity and transit planet
searches, the past 7~years have yielded several imaged Jovian planets between 10 and 70~AU separations, including four planets around HR 8799,\cite{marois2008science,marois2010} beta Pic b,\cite{lagrange2009} and HD
95086 b.\cite{rameau2013} Another half dozen low-mass ratio and/or wider separation planet-mass companions have also been imaged (e.g. NICI)).\cite{lafreniere2008pc,delorme2013,currie2014rox,Naud2014} The limited successes of large imaging campaigns even though hundreds of stars have been observed is hinting that massive Jovians planets are possibly rare $> 10$~AU around stars. \cite{lafreniere2007,nielson2008,chauvin2010,Nielsen2013,biller2013,vigan2012,chauvin2014} 

Direct exoplanet imaging is a challenging endeavour, as planets are located very close to their host star and are orders of magnitude fainter.  It was discovered that quasi-static speckles, i.e. noise in the stellar point spread function (PSF) halo coherent for several seconds to hours, are the most important limiting factor when trying to suppress the stellar halo of adaptive optics images.\cite{marois2003}  Several ideas to suppress the quasi-static speckles have been proposed and tested including: using the spectral differences between stars and planets by differencing polychromatic images (simultaneous spectral differential imaging SSDI),\cite{racine1999,marois2000,sparks2002} using the intrinsic field rotation on altitude/azimuth mount telescopes to decouple stationary quasi-static speckles from slowly revolving point sources (angular differential imaging, ADI),\cite{marois2006} using the polarization difference between the unpolarized stellar halo and the polarized light from a disk (PDI),\cite{Kuhn2001} or using the speckle coherence with the stellar light to discriminate between coherent speckles than incoherent point sources (CDI).\cite{codona2004}

While ADI is now being widely used in conjunction with the Locally Optimized Combination of Images (LOCI) algorithm (the LOCI algorithm consists of using a least-squares to optimaly combine a set of reference PSF images to subtract the quasi-static speckle noise), \cite{marois2006,lafreniere2007loci} SSDI has so far been struggling, with very limited successes. One of the main difficulties in implementing an SSDI PSF subtraction technique is non pupil-conjugated optical elements that generate PSF chromaticity therefore preventing a high level of speckle subtraction.\cite{marois2008} A new class of instruments designed to minimize chromaticity, e.g. through use of  very smooth optics, such as the Gemini Planet Imager (GPI) and SPHERE,\cite{macintosh2014,beuzit2008} are currently being used or commissioned on large 8-m class telescopes with the hope of finally allowing a high degree of SSDI speckle subtraction. These instrument include an integral field spectrograph (IFS) to simultaneously acquire many spectral channels. The second complexity with SSDI is the potential large variation in planetary flux as a function of wavelength, e.g. due to broad methane features. This variation can be an asset, allowing nearby wavelength channels to be compared without self-subtracting the planet, but needs careful attention to produce optimal results.  A first attempt to solve this problem has been presented by the P1640 team,\cite{pueyo2012} where the authors have used a damped-LOCI algorithm. In this paper, we are presenting a new algorithm, similar to damped-LOCI, but optimizes differently. The new algorithm aimed at subtracting PSF images using ADI, SSDI, PDI, CDI or archive-based PSFs, acquired with various ground- or space-based instruments, but optimize, at the same time, the planet signal-to-noise ratio (SNR) with a single least-squares step.

The paper layout is as follows. \S~\ref{sectssdi} describes the new TLOCI (for Template LOCI) algorithm while performance using GPI on-sky data is presented in \S~\ref{sectgpi}. The spectrum extraction accuracy is presented in \S~\ref{sectspec} and the technique to derive realistic error bars is discussed in \S~\ref{sectphot}. The performance difference between doing only ADI or only SSDI is presented in \S~\ref{sectadissdi}. Conclusions are presented in \S~\ref{conc}.

\section{Speckle Suppression with SSDI}
\label{sectssdi}
\subsection{The SSDI challenge}
The original LOCI algorithm was designed mainly for ADI observations where the planet moves continuously in position angle while having a constant flux, while neglecting sky transmission variations, PSF smearing due to fast field rotation and seeing/Strehl variations. For ground-based observations, ADI imaging consists of acquiring many short single band exposures for a typical hour-long sequence. For each image ($I_i$) where we want to subtract the quasi-static speckle noise, a reference image ($I^{\rm{REF}}$) is derived using a linear combination of the other images of the same sequence.

\begin{equation}
I^{\rm{REF}}_i = \sum_{\rm{k}} (c_{\rm{k,i}} \times I_{\rm{k}}).
\end{equation}

\noindent The various weights ($c_{\rm{k,i}}$) to give to each reference image are obtained by solving the following linear equation:

\begin{equation}
A c_i = b_i,
\end{equation}

\noindent where $A$ is the reference images covariance matrix, $c_i$ is the coefficients vector and $b_i$ is the correlation vector of the image to subtract with all reference images. The covariance matrix $A$ is inverted and matrix multiplied to $b_i$ to find the coefficient vector $c_i$ (i.e. LOCI is a least-squares fit). The LOCI algorithm is thus finding the set of coefficients such that the residual noise, after subtraction of the reference image, is minimized, thus greatly improving our ability to detect point sources. To avoid point source self-subtraction, for any given image, other images acquired close in time where point sources partially overlap (field rotation typically less than $\frac{1}{2}\lambda/\rm{D}$) are removed from the reference image set. Image subtraction is usually performed in small sections, typically annular sections, as the motion of point sources due to the field rotation between images is increasing with separation. The main advantage of doing this local optimization is to minimize the number of rejected images, thus maximizing the available information to subtract the PSF at each separation. 

The Speckle-Optimize Subtraction for Imaging Exolanets (SOSIE) pipeline\cite{marois2010sosie}  and similar approaches reported in other, later recent studies have adopted the general LOCI formalism but include upgrades to improve contrast and limit biases in the original LOCI algorithm.\cite{currie2012,currie2013} First, if the reference image archive is too large (and sometime well decorrelated with the image to subtract), noise can propagate into the matrix inversion that prevents an optimal set of coefficients to be found.\cite{marois2010sosie} To solve this problem, the covariance matrix can instead be inverted and truncated using a single value decomposition (SVD) and/or modified using a correlation-based image selection techniques.\cite{marois2010sosie,currie2012} These methods yield up a factor of 10 gain with {\it Hubble Space Telescope} NICMOS data and smaller but significant gains from ground-based conventional AO system data.\cite{currie2013}  A very similar algorithm (Karhunen-Lo\`{e}ve Image Projection (KLIP),\cite{soummer2012} was presented to achieve a similar optimization by decomposing the reference image archive into eigenvectors (for covariance matrices, doing an SVD or eigenvector decomposition are mathematically equivalent), so SOSIE and KLIP are essentially using the same least-squares matrix inversion algorithm scheme. Second, SOSIE algorithm is using a masking technique to prevent the LOCI algorithm to partially subtract the planet (using the speckle noise under the planet) in the process of trying to subtract the noise, greatly improving the planet algorithm throughput.\cite{marois2010sosie} The team also presented the idea of generating template PSFs to fit the astrometry and photometry of companions and allow contrast curve calculation without having to inject artificial sources.

Trying to extend this LOCI algorithm for polychromatic data suffers from four main problems:
\begin{enumerate}
\item Speckle position scales with wavelength, so images acquired at different wavelengths need to be magnified to align speckles. This magnification radially shifts true point sources. 
\item Two adjacent wavelength PSF images can be highly correlated, but significant point source overlap will result in significant self-subtraction.
\item Planets may exhibit large flux variations from wavelength-to-wavelength, like the methane absorption bandhead at 1.6~microns.
\item  The unkown planet spectrum is producing hard-to-calibrate self-subtractions, complicating the planet PSF forward-modeling and planet photometric/spectroscopic extractions.
\end{enumerate}

Early SSDI attempts used a limited number of polychromatic PSFs and calculated optimal single and double differences (SD and DD).\cite{marois2000} The DD is the curvature of the flux as a function of wavelength. If a planet is present, it will be generating a strong curvature relative to speckles that are mostly linear with wavelength. A second similar approach consisted in using many polychromatic images and fit a low order polynomial pixel per pixel.\cite{sparks2002} Although these approaches are valid, they do rely on expected speckle behaviour with wavelength. Using the NICI instrument where both ADI and SSDI are used (two wavelengths are acquired simultaneously with the instrument), the ADI and SSDI techniques were combined by simply adding the extra wavelength to the LOCI ADI reference image set.\cite{artigau2008}

The damped-LOCI algorithm\cite{pueyo2012} introduced a least-squares fit for solving the many-wavelengths SSDI problem. Instead of blindly performing the covariance matrix inversion, a penalty term is added to force the algorithm to find positive coefficients and to maximize the flux in a known $\sim \lambda$/D aperture where a known planet is located. Once a standard LOCI algorithm has found a planet, to properly retrieve its spectrum, a small aperture is placed on the companion and the damped-LOCI algorithm is run to try to avoid local contamination. The main drawback of this approach is that the location of the planet needs to be known in advance (usually not the case), and while trying to maximize the flux, the algorithm could be maximizing the speckle noise under the planet, thus contaminating the planet flux by an unknown amount.

   \begin{figure}
   \begin{center}
   \begin{tabular}{c}
   \includegraphics[height=10cm]{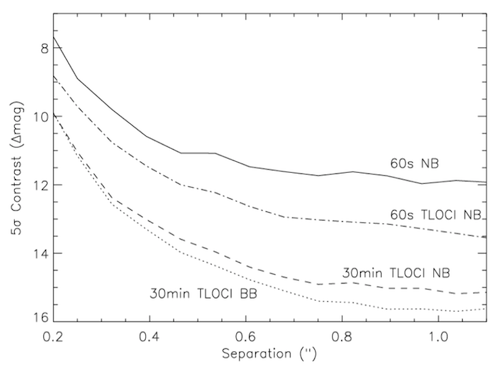}
   \end{tabular}
   \end{center}
   \caption[example] 
   { \label{TLOCIcontrast} 
The GPI bright star (H = 3.54; Beta Pic acquired on Nov 18, 2013) $5\sigma$ contrast for a 60s exposure in a single data cube slice (NB) after applying an unsharp mask to remove the seeing halo (solid line) and after TLOCI (a combination of ADI and SSDI image subtraction using an estimated spectrum for the planet - here DUSTY\cite{baraffe2002}) PSF subtraction (dot-dashed line). The dashed and dotted lines are showing the contrast for a 30 minutes sequence (30x60s data cubes) after PSF subtraction using the TLOCI algorithm for a single slice (NB, dashed line) and for a broadband image (BB, collapse data cube; dotted line). For this sequence the average air mass and DIMM seeing were 1.08 and 0.68~arcsec respectively in visual band. The GPI cryocoolers were set to minimum power during the sequence to reduce vibration.}
   \end{figure} 

\subsection{First verison of TLOCI}

The TLOCI SSDI speckle subtraction algorithm differs from the original LOCI (minimizing the noise) or damped-LOCI (minimizing the noise while trying to maximize the flux in a aperture where a known planet is located), by trying to maximize the SNR of any point sources in an image.\cite{marois2014tloci} To achieve this optimization, the reference image covariance matrix $A$ is penalized by adding an additional covariance matrix $A^{\rm{REF}}$ weighted by the parameter $\Lambda$:


\begin{equation}
    A^\prime = (1 - \Lambda) A + \Lambda \times A^{\rm{PSF}},
\end{equation}

While this approach was promising for small penalties, we encountered problems when trying to force the algorithm in extreme cases, i.e. having a large weight on the $A^{\rm{PSF}}$ matrix for highly accurate planet characterization. The algorithm failed to converge to a reference image that would conserve the planet flux. We therefore developed a variation on the algorithm. 

   \begin{figure}
   \begin{center}
   \begin{tabular}{c}
   \includegraphics[height=8cm]{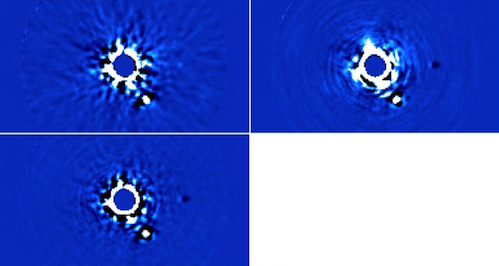}
   \end{tabular}
   \end{center}
   \caption[example] 
   { \label{TLOCIimvaralgo} 
TLOCI PSF subtraction on Beta Pic. Upper left: ADI only, upper right: SSDI only, bottom left: ADI+SSDI. The TLOCI spectral coefficients were calculated assuming a DUSTY (smooth-spectrum) planetary PSF for SSDI and ADI+SSDI and to collapse the data cube into a final image. Note the azimuthal negative side-lobes for ADI, radial negative side lobes for SSDI and circular negative side lobes for ADI+SSDI. A reference image selection criteria of 50\% was chosen for all three cases. A $\Delta$mag$~= 12.5$ methane companion was added at 0.63$^{\prime \prime}$ West of the star; even though the companion does not match the template it is easily detected, albeight at a lower SNR.}
   \end{figure} 

\subsection{Second version of TLOCI}

A second, simpler (both in concept and implementation) version of the TLOCI algorithm was developed for the Gemini Planet Imager. While not being as theoretically optimal in its search for the coefficients that maximize SNR, it does a proper combination of the ADI and SSDI images as well as reducing in pathalogical cases to a single step least-squares noise subtraction process. The new algorithm goes back to simply selecting images and rejecting others to avoid strong planet flux contamination. Instead of blindly removing some images due to some minimal field rotation as the original LOCI, the new TLOCI code keeps images on the basis of their potential contamination of the planet flux. With the user-defined template planet spectrum, TLOCI uses a single parameter to determine the degree of acceptable self-subtraction. As in the damped-LOCI algorithm, only positive coefficients are used. This avoids large countervailing positive and negative coefficients in regions of the PSF where degeneracies exist, but in the process, can generate complex planet flux contamination if the planet is strongly present in one image and not in the other. Before calculating the coefficients $c_{i,k}$, all images are normalized so that they all have the same total flux. Using positive coefficients implies that all coefficients are between 0 and 1. Thus, the maximum weight that a reference image can have is a coefficient of 1. This happens if all other coefficients are 0. As a consequence, if the potential contamination of the planet flux in a reference image is 20\%, we know from positivity that the maximum contamination caused by this reference in $I_i^{\rm{REF}}$ is 20\%. Since the sum of all the coefficients is 1, if we choose a 20\% contamination threshold for selecting the images used as references, we know that the maximum amount of contamination in $I_i^{\rm{REF}}$ is 20\%. The algorithm steps are as follows (for a typical GPI sequence and data cube images):

\begin{enumerate}
\item Images are first registered, are then spatially magnified (wavelength by wavelength) to align speckles, and are finally flux normalized to flatten the spectrum of the star. In the case of GPI, this is done using the four calibration satellite spots.\cite{wang2014}
\item The satellite spots of all datacube slices of all the images of the sequence are combined to generate a master PSF.
\item A planet spectrum is selected by the user to optimize the PSF subtraction. That template spectrum could be a flat spectrum, a methane-dominated spectrum, a DUSTY spectrum\cite{baraffe2002} or any other used-defined spectrum. 
\item The image data cubes are then divided into annuli, and the template PSF is used to simulate an artificial planet in those annuli. The simulated planet is moved in position angle to follow the field-of-view rotation and it is moved radially to account for the magnification that was applied to align speckles. At each wavelength, the planet flux is normalized following the user-selected planet spectrum.
\item In the annulus we want to subtract, an aperture is used over the artificial planet to integrate the planet flux. The same aperture is then used to integrate the artificial planet flux in the reference images (where the artificial planet is located at a different position and may have a different intensity), and the ones that contain less than a user-defined flux contamination value are kept to run the least-squares.
\item Once the appropriate reference images have been selected, the least-squares is run using positive coefficients and SVD cut-off to generate a reference annulus image. That reference image is then subtracted. 
\item The same coefficients are then applied to the artificial planet images to determine the subtracted planet PSF and to estimate a throughput flux correction that will be applied to produce a throughput-corrected final combined image.
\item The steps are repeated for all annuli of an image, all wavelengths and all images of the sequence.
\item Each subtracted data cube is then rotated to have North up and median combined.
\item The final data cube is then collapsed into a final 2D image by using the expected planet spectrum and the noise in each slice to perform a weighted mean.
\item The overall steps can be repeated using a different planet flux conservation number, a different SVD cut-off or different kind of planet spectrum to optimize the SNR and search for specific type of planets (i.e. DUSTY or methane planets of different temperatures).
\end{enumerate}

   \begin{figure}[h]
   \begin{center}
   \begin{tabular}{c}
   \includegraphics[height=9cm]{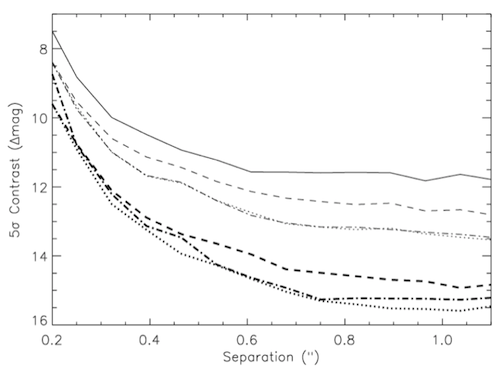}
   \end{tabular}
   \end{center}
   \caption[example] 
   { \label{TLOCcontsubalgo} 
The Beta Pic (November 18, 2013, H-band) sequence, with various PSF subtraction algorithms. The solid line is the 60s raw contrast, while the thin dashed, dot-dashed and dotted lines are the TLOCI ADI, SSDI and SSDI+ADI PSF subtraction residuals respectively (the DUSTY spectrum is used here for the SSDI and SSDI+ADI cases). The three thick dashed, dot-dashed and dotted lines are the 30 minutes combined curves (ADI, SSDI, and ADI+SSDI). While some gain can be achieved with ADI and a least-squares, a better contrast, up to a magnitude, can be achieved with SSDI and SSDI+ADI. TLOCI PSF subtraction that combines both SSDI and ADI into a single algorithm achieves the best contrast overall, being slightly better than using SSDI alone. The SSDI and SSDI+ADI gain would be even higher for methane-like spectra (strong spectral features).}
   \end{figure} 

In contrast with the damped-LOCI algorithm, the new TLOCI algorithm can be used to maximize the detection of a given template spectrum no matter where in the image the planet is located, i.e. the planet location does not need to be known in advance. This technique can thus discover new planets by finding a set of coefficients that are optimal for a specific planet spectrum of interest. The TLOCI algorithm is also working at preventing flux contamination of images where important point source overlap is present, even in cases where the planet flux is significantly changing between wavelength images. By adjusting the subtraction aggressiveness, the users can optimize the final SNR by finding a compromise between subtracting the noise and conserving the planet flux.

   \begin{figure}
   \begin{center}
   \begin{tabular}{c}
   \includegraphics[height=9cm]{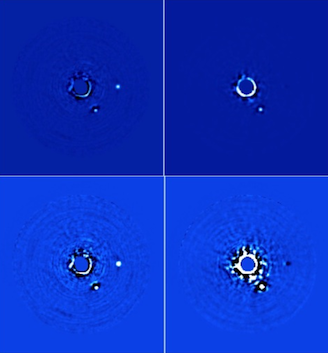}
   \end{tabular}
   \end{center}
   \caption[example] 
   { \label{TLOCIimvarspec} 
TLOCI (ADI+SSDI) PSF subtraction after combining a 30 min sequence (Beta Pic H-band Nov 18, 2013). A $\Delta$H $= 12.5$ artificial methane point source is added 0.63$^{\prime \prime}$ West of the star. Left panels: TLOCI optimized for methane objects. Right panels: TLOCI optimized to find dusty planets. The bottom and upper panels identical to the above images but with a different display range.}
   \end{figure} 

The TLOCI algorithm can also be applied on single band ADI sequences, on ADI and SSDI sequences (and PDI if such data is also available) to do the subtraction as a single speckle subtraction step, use reference images obtained using CDI (and perform ADI and SSDI at the same time), or use additional reference images from an image archive.  

\section{TLOCI GPI testing}
\label{sectgpi}
The new TOCI algorithm has been tested using GPI commissioning H-band data of Beta Pictoris acquired November 18, 2013 (first light run). A total of thirty 60s exposures near transit were acquired in good seeing conditions and extracted using the GPI data reduction pipeline.\cite{marshall2014} The image reduction and the TLOCI PSF subtraction was performed using a standalone IDL software (it is a goal to bring the main TLOCI algorithm in the GPI data reduction pipeline in the near future). The images were filtered using a soft unsharp mask ($5 \times 5 \lambda$/D) to remove the low spatial frequencies (this ensure that the least-squares is working at subtraction the $\lambda$/D noise instead of the low spatial frequencies) and then spatially magnified (using the image header wavelength information) to align speckles. A Gaussian was then fitted on the four satellite spots to find the star center and all slices of each images were shifted to the image center. Due to some non-repeatable flexure that sometime offsets the wavelength,\cite{schuyler2014} the spot locations were used to readjust the magnification to better overlap the spots and, in the process, the speckles. The images were then flux normalized to show the same total flux using the same total flux for the four spots. The TLOCI algorithm as described in the previous section was then applied. The resulting contrast curve for a DUSTY spectrum is shown in Fig.~\ref{TLOCIcontrast}. The final combined image for various types of reduction (ADI only, SSDI only or SDSDI+ADI using the TLOCI algorithm) are shown in Fig~\ref{TLOCIimvaralgo}. The corresponding contrasts for these cases are shown in Fig.~\ref{TLOCcontsubalgo}. An artificial $\Delta$mag$ =12.5$ methane planet was added to the data West of the star and the sequence subtracted using a DUSTY and then a methane planet spectrum for the subtraction optimization (see Fig.~\ref{TLOCIimvarspec}).

\section{The challenges of spectral extraction}
\label{sectspec}
Once a real planet has been discovered with GPI, the next challenge is to extract its spectrum for atmospheric characterization. Given that TLOCI uses an a priory of what type of spectrum to search for in the data, the selected spectrum biases the recovered planet spectrum. The bias is extremely difficult to calibrate, leaving only less aggressive subtractions or even limit the TLOCI algorithm to only use ADI (thus removing the SSDI spectral prior) to subtract images.

This bias can be better understood by considering the following: consider that we are iterating around a good input spectrum to find the real planet spectrum using both ADI and SSDI with TLOCI. If the planet spectrum is different than the TLOCI input spectrum, this means the algorithm is not properly calculating the flux contamination between images, thus generating an unknown amount of flux contamination (proportional to the difference between the real planet spectrum and the user-defined spectrum) in the resulting combined data cube, even when using positive least-squares coefficients. The TLOCI template PSF used to extract the planet photometry and astrometry will also differ from the real planet PSF given that the algorithm is using the wrong spectrum, further adding uncertainty to the photometry and astrometry. More work is needed to better understand this bias and how it could be minimized or calibrated to extract higher-SNR planet spectra.


   \begin{figure}
   \begin{center}
   \begin{tabular}{c}
   \includegraphics[height=6cm]{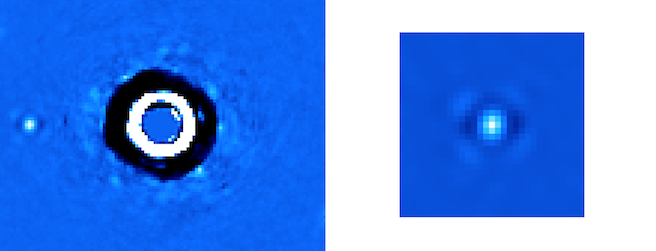}
   \end{tabular}
   \end{center}
   \caption[example] 
   { \label{HD172555psf} 
Left panel: The HD172555 PSF subtracted image with an artificial planet East of the star (the inner two annuli have not been contrast optimized). Right panel: The TLOCI estimated PSF.}
   \end{figure} 

\section{Determining Realistic Photometric and Astrometric Error Bars}
\label{sectphot}
Following various developments in the initial SOSIE code, similar techniques have been implemented in TLOCI for GPI data analysis. First, the algorithm uses a masking technique to avoid the least-squares algorithm subtracting the planet flux; planet flux is never part of the reference section to calculate the least-squares coefficients. The template PSF is then used to simulate the planet PSF after image subtraction (see Fig.~\ref{HD172555psf}) and to estimate the amount of self-subtraction to properly normalize the final image, thus showing a uniform planet throughput in each annulus and in the entire image (all annuli). Error bars are a combination of:

\begin{enumerate}
\item The background noise relative to the planet flux (the planet PSF is subtracted using the template PSF; the flux error is determined by adding similar artificial planets at various position angles and their flux extraction variation corresponds to the real PSF extraction error).
\item The PSF subtraction residual. PSF subtraction residuals that are larger than the background noise is a sign that the photometry and astrometry are dominated by differences between the real planet PSF and the template PSF.
\item The planet spectrum is obtained relative to the star spectrum. For GPI, the star spectrum is derived using the four satellite spots. The four spot SNRs in raw unsubtracted images are thus contributing to the final planet spectrum error. Note that the speckle noise is highly correlated between spectral channels in raw GPI images, so that noise is more a global offset in intensity than a wavelength-to-wavelength error.
\item Planet photon noise
\end{enumerate}

One key issue that requires careful attention is how the various noises are correlated between spectral channels as this can impact model fitting when performing $\chi^2$ analysis and wrongly assuming uncorrelated noise between spectral data points.

Another interesting effect was discovered when a planet is lost in the residual noise. The planet searching algorithm converged at finding a bright positive speckle in the search box instead of the planet, so it is finding, on average, a $\sim 1\sigma$ planet as a minimum planet flux (instead of $\sim $zero), even if there is no flux from the planet at that wavelength and the mean background flux is zero.  A simple solution to remove that flux estimate bias is to only do PSF fitting if the planet is convincingly detected (more than $3\sigma$) and define $3\sigma$ upper limits if it is not.

   \begin{figure}
   \begin{center}
   \begin{tabular}{c}
   \includegraphics[height=6cm]{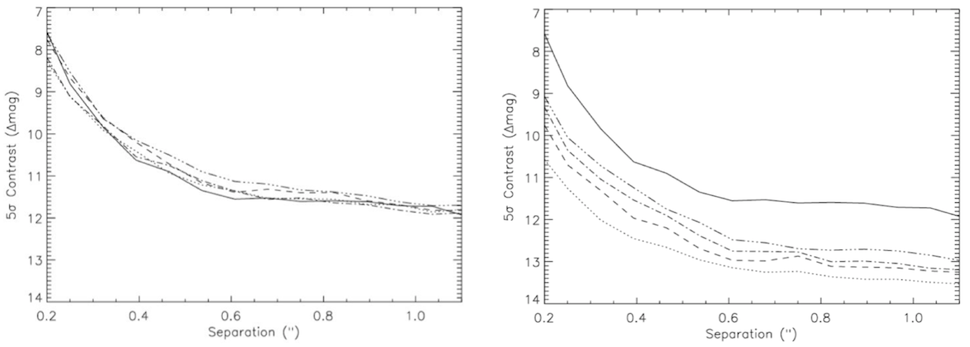}
   \end{tabular}
   \end{center}
   \caption[example] 
   { \label{TLOCIcorcontcomb} 
Left panel: PSF subtraction achieved by the subtraction of images acquired at the same wavelength, but at a different time (same object and sequence). The solid line is the raw H-band 60s contrast (Beta Pic), while the dotted, dashed, dot-dashed and triple dot-dashed lines are for the subtraction of slice 20 of various files (see below). The images are showing very little correlation between images, even for back-to-back images. Right panel: PSF subtraction achieved by the subtraction of images acquired simultaneously, but at a different wavelength. The solid line is the raw H-band 60s contrast (Beta Pic), while the dotted, dashed, dot-dashed and triple dot-dashed lines are for the subtraction of slice 20 with the slice 22, 26, 30 and 34 respectively. The images are currently much more correlated as a function of wavelength than as a function of time; some speckle evolution with wavelength is detected. This higher correlation confirms that SSDI-based PSF subtraction algorithms will perform significantly better than ADI-based codes at subtracting the PSF for GPI, especially when searching for objects having strong spectral features, such as methane planets.}
   \end{figure} 

\section{GPI ADI vs SSDI performances}
 \label{sectadissdi}
Visual inspection of the GPI first run data showed that it exhibits significant image-to-image speckle evolution for reasons not yet fully understood. This lack of PSF time stability does have implications as the ADI technique relies on that stability to suppress the quasi-static speckles. Looking at how image subtracts as a function of time and wavelength clearly shows this effect (see Fig.~\ref{TLOCIcorcontcomb}).

\begin{figure}
   \begin{center}
   \begin{tabular}{c}
   \includegraphics[height=8cm]{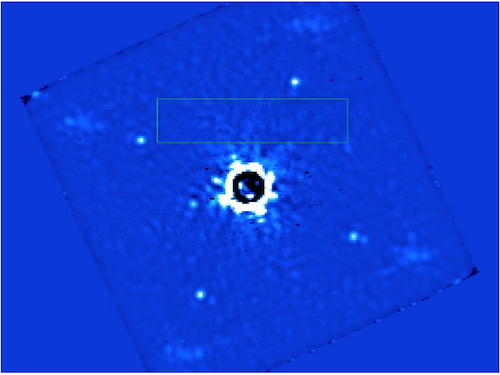}
   \end{tabular}
   \end{center}
   \caption[example] 
   { \label{TLOCIcorsection} 
A typical GPI image. The green box is the section being analyzed to evaluate the speckle correlation as a function of time and wavelength. The correlation analysis is performed on the magnified images, i.e. where the spots and speckles are spatially aligned. An unsharp mask has also been used to remove the low spatial frequencies and background flux.}
   \end{figure}

\begin{figure}
   \begin{center}
   \begin{tabular}{c}
   \includegraphics[height=6.5cm]{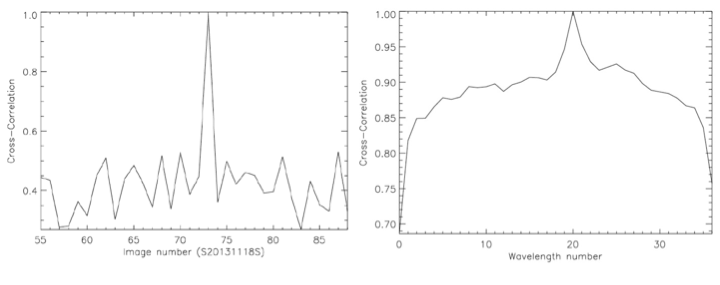}
   \end{tabular}
   \end{center}
   \caption[example] 
   { \label{TLOCIcorcontcomb2} 
Left panel: Speckle correlation at the same wavelength (slice 20) for a sequence of images (Beta Pic). The reference image is 73. Right panel: Speckle noise correlation between wavelength channels (Beta Pic, slice 20 is the reference). The strong peak near the reference slice is probably due to the pipeline generating spectraly oversampled highly-correlated images (we should have 15 independent spectral channels while the pipeline outputs 37 wavelength slices). When compared to the left panel (correlation as a function of time), speckles are much more correlated as a function of wavelength.}
   \end{figure}

To quantify the time and spectral speckle correlation, we take the same November 18, 2013 Beta Pic sequence and plot the auto-correlation function of an image section (see Fig.~\ref{TLOCIcorsection}). The resulting speckle auto-correlations are shown in Fig.~\ref{TLOCIcorcontcomb2}. The speckles are more than twice as correlated as a function of wavelength than as a function of time, confirming our initial visual inspection. Finding the cause of that time instability and solving it is important, especially for characterizing planets where SSDI needs to be used with caution.

\section{Conclusions}
\label{conc}
A new ADI and SSDI PSF subtraction algorithm (TLOCI) was presented to optimally combine a set of polychromatic images to maximize the speckle noise subtraction, while ensuring minimum planet flux contamination from point source self-subtraction. The algorithm can be used to maximize the signal-to-noise ratio of point sources having a specific user-defined spectrum; in contrast to older algorithms that were trying to minimize the noise, without any consideration of how the planet flux would be affected by such choice of least-squares coefficients. 

The new algorithm is maximizing the detection of planets using an input spectrum as an input parameter and ensures a reliable spectrum extraction of detected planets if the aggressiveness of the subtraction is scaled back to minimize point source overlap and the use of SSDI. The ADI-only less aggressive extracted spectrum is consistent with the image noise, showing that they were properly retrieved. The use of non-optimal input spectra and aggressive subtractions do produce a complicated bias. While the speckles are highly correlated as a function of wavelength, the lack of good time stability in the GPI data means that longer sequences will be required to properly characterize faint planets.

The relatively good contrast obtained for the methane-like planets is good news for the GPI campaign. The high level of speckle rejection is mainly due to the careful design of GPI optical layout that minimizes chromatic Fresnel propagation effects and high quality optics. The performance is not as good for a flat (or DUSTY) spectrum objects, but such spectra are more typical of background stars and warm and bright planets; these planets are easier to detect and thus suffer less from the less optimal contrast. 

Future algorithm developments will be geared toward designing a PSF archive to add more reference images, this could potentially help improve the contrast at smaller separations. Given the lack of time stability, it is unclear if this approach will help improve the contrast or not. Another aspect that will be explored is how to optimize TLOCI for disk imaging.

This new, more optimal planet finding algorithm can be used for ADI, PDI, CDI and polychromatic data (all these techniques performed as a single speckle subtraction step) will play a crucial role in maximizing the science discoveries of GPI and other similar instruments, and will be an important asset in the search of Earth-like planets with space coronagraph missions and extremely large ground-based telescopes. 

\subsection{Acknowledgments} 
Based on observations obtained at the Gemini Observatory, which is operated by the 
Association of Universities for Research in Astronomy, Inc., under a cooperative agreement 
with the NSF on behalf of the Gemini partnership: the National Science Foundation 
(United States), the National Research Council (Canada), CONICYT (Chile), the Australian 
Research Council (Australia), Minist\'{e}rio da Ci\^{e}ncia, Tecnologia e Inova\c{c}\~{a}o 
(Brazil) and Ministerio de Ciencia, Tecnolog\'{i}a e Innovaci\'{o}n Productiva (Argentina). C. Correia acknowledges the support of the European Research Council through the Marie Curie Intra-European Fellowship with reference FP7-PEOPLE-2011-IEF, number 300162.


\bibliography{report}   
\bibliographystyle{spiebib}   

\end{document}